\documentstyle[preprint,aps]{revtex}
\begin{document}
\preprint{UM-P-95/99, RCHEP-95/23}
\draft
\title{HYBRID $SU(2) \times U(1)$ MODELS, ELECTRIC CHARGE
NONCONSERVATION AND THE PHOTON MASS}
\author{A.Yu.Ignatiev\cite{byline1} and G.C.Joshi\cite{byline2}}
\address{Research Centre for High Energy Physics, School of
Physics, University of Melbourne, Parkville, 3052, Victoria,
Australia}
\maketitle
\begin{abstract}
Hybrid $SU(2) \times U(1)$ models are the models in which 
$SU(2) \times U(1)$ symmetry is broken down 
not only spontaneously (as
in the Standard Model), but also
explicitely by adding a hard mass term for the $U(1)$ field in 
the lagrangian.  We study the issue of electric charge 
nonconservation
and dequantization in these models. For this purpose we
construct and analyze a series of hybrid models with different
scalar contents.
We show that some of these models posess an interesting property:
the photon can remain massless (at least, at the tree level) 
even though the electric charge is not conserved.
\end{abstract}

\section{Introduction}
The electric charge conservation and the masslessness of photon are
the two fundamental ingredients of the Standard Model.  They have been
tested experimentally many times with high precision and at present we
have no evidence whatsoever that would question their validity.

Yet, over the past two decades there has been considerable interest in
constructing and analysing models in which one or both of the above
postulates do not hold true 
\cite{GN,we1,okun,we2,s,ts,Qnoncons,mtt,mn}. 
The purpose of these works was to
understand better the reasons why the electric charge is conserved and
the photon is massless. Also, without these works it would be very
hard to think of new experimental ways of testing these laws.

One of the important discoveries made in those works was the close
relation between the two ideas: electric charge (non)conservation and
electric charge (de)quantization \cite{we1,we2,Mel}.

All previous works on the subject have shared one common property: it
was impossible to violate the electric charge conservation without
giving the photon a (tiny) mass at the same time. This is not
surprising at all if we consider Maxwell equations of classical
electrodynamics. Since observational limit on the photon mass is
very tight (it has to be less than $ 10^{-24} $ GeV or even 
$ 10^{-36} $ GeV \cite{partdata}), model-building is severely 
restricted.

In this paper, our question is: can we make a model without that
undesirable property? In other words, can one modify the Standard
Model in such a way that the electric charge is not conserved but the
photon is exactly massless? Based on all our previous experience with
that kind of models, the answer would seem almost definitely, no.

However, in the present work we show how to construct a model in which
the electric charge is not conserved but the photon {\em is massless
at the tree level \/}.\footnote{The important further questions if 
this
property survives at higher orders of the perturbation theory and if
the model is renormalizable, are left open.}

Briefly speaking, the main idea is as follows. We point out that the
photon mass in the most general case gets two contributions: the first
from the mass term of the U(1) field (usually denoted by B) and the
second from the spontaneous breaking of the electromagnetic symmetry.
By the appropriate choice of parameters we can ensure that these two
contributions cancel each other so that the photon remains massless
(at least, at the tree level) despite the fact that the electric
charge is not conserved.

The plan of the work is this: in Section 2 we reproduce some familiar
formulas of the Standard Model in order to establish our notation and
to make easier the comparison with the further material. Section 3
deals with the case when the Standard Model is extended by adding a
hard mass term for the U(1) gauge field. In Section 4 we add to the
model of Section 3 an electrically charged scalar singlet with non-
zero vacuum expectation thus violating spontaneously the conservation
of electric charge. In Section 5 we substitute the scalar singlet by
the scalar doublet (with electric charge violating vacuum average)
leaving the rest of the model the same as in Section 4. Our discussion
and conclusions are contained in Section 6.

\section{Standard Model} 

We shall restrict ourselves to only one generation of quarks and
leptons (the addition of other generations will only clutter the
notation without giving any new insights).

The part of the total Lagrangian which is relevant for our purposes is
this
\begin{eqnarray}
{\cal L}_0 &=& {\cal L}_l + {\cal L}_q + {\cal L}_s\\
{\cal L}_l &=& \bar{L} (i \partial + g {\tau^a \over 2} A^a -{g' \over
2} B) L + \bar{e}_R (i \partial -g' B) e_R,\\
{\cal L}_q &=& \bar{Q} (i \partial + g {\tau^a \over 2} A^a +{g' \over
6} B) Q + \bar{u}_R (i \partial +{2g' \over 3} B) u_R + \bar{d}_R (i
\partial -{g' \over 3} B) d_R,\\
{\cal L}_s &=& |(\partial_{\mu} -ig {\tau^a \over 2} A^a_{\mu} -i {g'
\over 2} B_{\mu}) \phi|^2 - h(\phi^{\dagger} \phi - { 1 \over 2}
v^2)^2.
\end{eqnarray}
where $A^a= A_{\mu}^a \gamma_{\mu}$ and $B^a= B_{\mu}\gamma_{\mu}$ are 
$SU(2)$ and $U(1)$ gauge fields,
\begin{equation}
 L =\left( \begin{array}{c}
\nu_{eL} \\ e_L
\end{array} \right) ,\;\;
 Q =\left( \begin{array}{c}
u_{L} \\ d_L
\end{array} \right) ,\;\;
 \phi =\left( \begin{array}{c}
\phi^+ \\ \phi^0
\end{array} \right) 
\end{equation}
After spontaneous symmetry breaking,
\begin{equation}
 \langle \phi \rangle = {1 \over \sqrt{2}} \left( \begin{array}{c}
0  \\ v
\end{array} \right) ,
\end{equation}
we are interested in the part of Lagrangian $L_0$ which is quadratic
in the gauge fields $A_i$ and $B$:
\begin{equation}
{\cal L}^{quad}_0 = {1 \over 8} v^2
[g^2(A^2_{1\mu}+A^2{2\mu}+A^2{3\mu}) + g'^2 B^2_{\mu} - 2 gg' A_{3\mu}
B^{\mu}].
\end{equation}

This gives us the following mass matrix for the pair of fields $A_3$
and $B$ :
\begin{equation}
 M^0 = \left( \begin{array}{cc}
M^0_{33} & M^0_{34} \\
M^0_{43} & M^0_{44}
\end{array}
\right) = \left( \begin{array}{cc}
{1 \over 4} g^2 v^2 & -{1 \over 4}gg' v^2 \\
-{1 \over 4}gg'v^2 & {1 \over 4} g'^2 v^2
\end{array}
\right).
\end{equation}

Diagonalizing this mass matrix we arrive at a pair of physical fields,
$A$ and $Z$ which are identified with the photon and Z-boson:
\begin{eqnarray}
A^3_{\mu} &=& Z_{\mu} \cos \theta + A_{\mu} \sin \theta \label{9}\\
B_{\mu} &=& A_{\mu} \cos \theta  - Z_{\mu} \sin \theta. \label{10}
\end{eqnarray}
Here, the Weinberg angle is given by the standard expression:
\begin{equation}
\sin ^2 \theta = {g'^2 \over g^2 + g'^2}.
\end{equation}
Now, changing the fields $A_3, B$ into $A, Z$ in our initial
Lagrangian we finally obtain the electromagnetic interactions of
quarks and leptons:
\begin{eqnarray}
{\cal L}^{em} &=& {\cal L}_l^{em} + {\cal L}_q^{em} \label{12}\\
{\cal L}_l^{em} &=& A_{\mu}[{1 \over 2}(g\sin\theta - g'\cos\theta)
\bar{\nu}_{L}\gamma^{\mu}{\nu}_L
-{1 \over 2}(g\sin\theta + g'\cos\theta) \bar{e}_{L}\gamma^{\mu}{e}_L 
\nonumber\\
&&- g'\cos\theta \bar{e}_{R}\gamma^{\mu}{e}_R] \label{13} \\
{\cal L}_q^{em} &=& A_{\mu}[{1 \over 2}
g\sin\theta(\bar{u}_{L}\gamma^{\mu}{u}_L -
\bar{d}_{L}\gamma^{\mu}{d}_L) + g'\cos\theta({1 \over
6}\bar{u}_{L}\gamma^{\mu}{u}_L + \nonumber\\
&&{1 \over
6}\bar{d}_{L}\gamma^{\mu}{d}_L +{2 \over
3}\bar{u}_{R}\gamma^{\mu}{u}_R - {1 \over
3}\bar{d}_{R}\gamma^{\mu}{d}_R)] \label{14}
\end{eqnarray}
From this formula we can read off the values of the electric charges
of quarks and leptons:
\begin{eqnarray}
Q_{\nu} &=& {1 \over 4}(g\sin\theta - g'\cos\theta)\\
Q_e &=& -{1 \over 4}g\sin\theta -{3 \over 4}g'\cos\theta\\
Q_e^5 &=& -{1 \over 4}(g\sin\theta - g'\cos\theta)\\
Q_u &=& {1 \over 4}g\sin\theta +{5 \over 12}g'\cos\theta\\
Q_u^5 &=& {1 \over 4}g\sin\theta - {1 \over 4}g'\cos\theta\\
Q_d &=& -{1 \over 4}g\sin\theta - {1 \over 12}g'\cos\theta\\
Q_d^5 &=& -{1 \over 4}g\sin\theta + {1 \over 4}g'\cos\theta
\end{eqnarray}
Consequently, the electric charges of neutron and proton are:
\begin{eqnarray}
Q_n &=& Q_u+2Q_d= - {1 \over 4}(g\sin\theta - g'\cos\theta)\\
Q_p &=& 2Q_u+Q_d={1 \over 4}g\sin\theta + {3 \over 4}g'\cos\theta.
\label{23}
\end{eqnarray}

At this point one may wonder why Eq.(\ref{13}) to (\ref{23}) 
do not look very
familiar. The reason is this: in writing down Eq. (\ref{13}) to 
(\ref{23})  
we have not taken into account the formula
\begin{equation}
\label{24}
g \sin \theta = g' \cos \theta,
\end{equation}

which holds true in the Standard Model. We did not use this equation
when deriving Eq.(\ref{13}) to (\ref{23})  because there exists a 
crucial
distinction between them: Eq.(\ref{24})  will {\em not apply} in the
extended models to be considered below (Sections 3, 4, 5) whereas Eq.
(\ref{13}) to (\ref{23}) {\em will still be true} in all those models 
(provided one
puts in the modified value for $\sin \theta$, see below.)

Of course, if one wants to stay within the Standard Model, than one
has to put   $g \sin \theta = g' \cos \theta$ in Eq.(\ref{13}) to 
(\ref{23}) to recover the standard form for the electromagnetic 
Lagrangian:
\begin{equation}
{\cal L}_{em}= A_{\mu} \sum_f Q_f \bar{f} \gamma^{\mu}f
\end{equation}
with the correct values of fermion charges $Q_f$:
\begin{equation}
Q_{\nu}=0, \;\; Q_e=-e, \;\; Q_u={2 \over 3}e, \;\; Q_d=-{1 \over 3}e.
\end{equation}
Needless to say, all axial charges $Q^{5}_{i}$ vanish identically in
the Standard Model.

\section{Minimal hybrid $SU(2) \times U(1)$ model} 

Let us consider a model which differs from the Standard Model only in
one point: its lagrangian contains a mass term for the $U(1)$ gauge
field $B$ (before spontaneous symmetry breaking):
\begin{equation}
{\cal L}'={\cal L}_0 +{1 \over 2}m^2 B^2_{\mu}.
\end{equation}
After symmetry breaking, we obtain from this lagrangian the following
mass matrix for the gauge fields:
\begin{equation}
 M' = M^0 + \Delta M' = \left( \begin{array}{cc}
M^0_{33} & M^0_{34} \\
M^0_{43} & M^0_{44} + m^2
\end{array}
\right) = \left( \begin{array}{cc}
{1 \over 4} g^2 v^2 & -{1 \over 4}gg' v^2 \\
-{1 \over 4}gg'v^2 & {1 \over 4} g'^2 v^2 + m^2
\end{array}
\right).
\end{equation}

Following the same path as in the standard case (Section 2), we
diagonalize the mass matrix to obtain the physical fields. Although
these fields are different from the standard fields (\ref{9}), 
(\ref{10}),
we keep the same notation for them: $A$ and $Z$ since we have to
identify them with the observable particles: photon and Z-boson:
\begin{eqnarray}
A^3_{\mu} &=& Z_{\mu} \cos \theta ' + A_{\mu} \sin \theta ' \\
B_{\mu} &=& A_{\mu} \cos \theta '  - Z_{\mu} \sin \theta '.
\end{eqnarray}

The gauge boson masses acquire small corrections (assuming that $m$
is small); in particular, the photon gets non-zero mass:
\begin{eqnarray}
M_Z &=& {1 \over 2} \sqrt{g^2+g'^2} v + O({m^2 \over v})\\
M_{\gamma} &=& gm + O({m^2 \over v}). \label{32}
\end{eqnarray}

The Weinberg angle gets a small correction, too (to avoid
misunderstanding, we note that we are working on the tree level
throughout the paper so the word "correction" has nothing to do with
perturbation theory):
\begin{equation}
\sin^2\theta ' =\sin^2 \theta + {m^2 \over M_Z^2} (1 - {e^2 \over 
\sin^2
\theta} + e^2 - \sin^4 \theta) \approx  0.64  {m^2 \over M_Z^2}.
\end{equation}

This fact leads to a drastic consequence: the electric charge non-
conservation. To show that, let us find the electromagnetic part of
the lagrangian.

If we compare the ways of reasoning in Sections 2 and 3 we shall see
that the same formulas, Eq. (\ref{13}) and (\ref{14})  
applies also in the present case;
the only change that should be made is to change $\sin \theta$ to
$\sin \theta^{'}$, the rest of the formulas being unchanged:
\begin{eqnarray}
{\cal L}^{'em} &=& {\cal L}_l^{'em} + {\cal L}_q^{'em} \\
{\cal L}_l^{'em} &=& A_{\mu}[{1 \over 2}(g\sin\theta' - 
 g'\cos\theta'
)\bar{\nu}_{L}\gamma^{\mu}{\nu}_L
-{1 \over 2}(g\sin\theta' + g'\cos\theta' )
\bar{e}_{L}\gamma^{\mu}{e}_L \nonumber\\
&&- g'\cos\theta' \bar{e}_{R}\gamma^{\mu}{e}_R] \label{35} \\
{\cal L}_q^{'em} &=& A_{\mu}[{1 \over 2}
g\sin\theta'(\bar{u}_{L}\gamma^{\mu}{u}_L -
\bar{d}_{L}\gamma^{\mu}{d}_L) + g'\cos\theta' ({1 \over
6}\bar{u}_{L}\gamma^{\mu}{u}_L +{1 \over
6}\bar{d}_{L}\gamma^{\mu}{d}_L + \nonumber\\
&& {2 \over
3}\bar{u}_{R}\gamma^{\mu}{u}_R - {1 \over
3}\bar{d}_{R}\gamma^{\mu}{d}_R)] \label{36}
\end{eqnarray}

Based on this formula, we can arrive at an important conclusion: as
soon as the equality $g \sin \theta = g' \cos \theta$  is broken, the
electromagnetic current conservation is violated immediately.
 To avoid confusion, one essential point needs to be emphasized here.
We have defined the electromagnetic current (and thereby the electric
charge) as the current interacting with (i.e. standing in front of)
the electromagnetic field $A_{\mu}$. Naturally, one can ask about the
standard fermion electromagnetic current of the form
\begin{equation}
j_{\mu}= e(-\bar{e} \gamma_{\mu}e + {2 \over 3} \bar{u} \gamma_{\mu}u
- {1 \over 3} \bar{d} \gamma_{\mu}d). \label{37}
\end{equation}

 Although this current is still conserved in the present model
, it  unfortunately becomes devoid of physical meaning,
because all physical processes and experiments are based on the
interaction between the charges and electromagnetic fields; therefore
in the framework of the present model we have to attach physical
meaning and reserve the name "electromagnetic current" for the current
of Eq.(\ref{35}) and (\ref{36}), rather than that of Eq. (\ref{37})

To summarise, this theory features three fundamental deviations from
the Standard Model: massiveness of photon, the electric charge
dequantization, and the electric charge non-conservation.

Now, let us discuss the experimental limits on the parameter $m$
which result from the above three features.

In our case, the experimental upper bound on the photon mass gives, by
far, the strongest constraint on the value of $m$. It has been
established that the photon mass should be less than $ 10^{-24} $  
GeV or even
$ 10^{-36} $ \cite{partdata}. Therefore, from Eq. (\ref{32}) we 
find that 
the parameter $m$ cannot
exceed $2 \times 10^{-24}$ GeV or $2 \times 10^{-36}$ GeV.   
With such small values of the parameter $m$,
the charge dequantization and charge non-conservation effects are
expected to be too small to be observed. For instance, the best
experimental limits on electric charge dequantization are given by the
following figures: neutron charge: $Q_n < 10^{-21}$ \cite{n};  
charge of an atom:
$Q_a < 10^{-18}$ \cite{ep}
neutrino charge: $Q_{\nu} < 10^{-13}$ \cite{brf} or $10^{-17}$ 
\cite{bc} 
(for a detailed discussion of these and other constraints, see 
\cite{bv}).

Thus we can conclude that the upper bound on the parameter $m$ imposed
by the masslessness of photon makes all other predictions very hard to
observe which  limits our interest in this model.

Note that this model (with no fermions) was first suggested in Ref. \
cite{clt} under the name
of "hybrid model". The authors of Ref. \cite{clt} were motivated by 
the
systematic search for renormalizable gauge models beyond the standard
$SU(2) \times U(1)$ model. As concerns the renormalizability of the
model which is certainly a very important issue, it has been proved in
Ref. \cite{clt}  that the theory posesses the property called tree 
unitarity
which is a weaker property than renormalizability. We are not aware of
any work which would further address the problem of renormalizability
of this type of models. Although it may appear to be of academical
rather than phenomenological character, this work would certainly be
very desirable because it would include or exclude  a whole new class
of gauge models from the set of renormalizable gauge theories. (Note
that we do not share the belief that non-renormalizability of a theory
automatically makes it physically uninteresting.)

\section{Hybrid model with a scalar singlet}

Let us now add to the Lagrangian of the previous section a piece
containing the scalar singlet field $\phi_1$ with the electric charge
$\epsilon$ (which coincides with the hypercharge in this case):
\begin{equation}
{\cal L}_1 = {\cal L}_0 + {1 \over 2}m^2B^2 + |(\partial_{\mu}  -i {g'
\over 2} B_{\mu}) \epsilon_1 ) \phi_1|^2 + P(\phi_1 , \phi). 
\end{equation}
Now, assume that the field $\phi_1$  has non-zero vacuum expectation
value $v_1$: $\langle \phi_1 \rangle = v_1$. Then, after spontaneous
symmetry breaking the mass matrix of the system $A_3, B$ is:
\begin{equation}
 M^1 = M^0 + \Delta M^1 = \left( \begin{array}{cc}
M^0_{33} & M^0_{34} \\
M^0_{43} & M^0_{44} + m^2 +{1 \over 2} g'^2 v_1^2 \epsilon_1^2
\end{array}
\right) = \left( \begin{array}{cc}
{1 \over 4} g^2 v^2 & -{1 \over 4}gg' v^2 \\
-{1 \over 4}gg'v^2 & {1 \over 4} g'^2 v^2 + m^2 +{1 \over 2} g'^2 
v_1^2
 \epsilon_1^2
\end{array}
\right).
\end{equation}
Performing the diagonalization as before, we obtain the mass of the
physical photon to be:
\begin{equation}
M^2_{\gamma} = g^2 (m^2 + {1 \over 2} g'^2 v_1^2
 \epsilon_1^2 ) 
\end{equation}

The formula for the photon mass (squared) consists of two
contributions: the first is proportional to $m^2$ (``hard mass'') and
the second is proportional to $v_1^2$ (``soft mass''). Nothing seems
to prevent us from considering {\em negative} values either for $m^2$
or for $v_1^2$. Thus we are led to a very interesting possibility: to
choose these two parameters in such a way that they exactly cancel
each other so that the photon remains massless\footnote{Here, we 
disregard a possible appearence of a Nambu-Goldstone boson. One may 
expect that its manifestations would be sufficiently suppressed, but
even if they were not, the model could be modified in analogy with
Ref. \cite{mtt}.}
(at least, at the tree
level):
\begin{equation}
m^2 + {1 \over 2} g^{'2} v_1^2 \epsilon_1^2 = 0. \label{41}
\end{equation}

Note that if this condition is satisfied, the Z-boson mass becomes
exactly equal to that of the standard Z-boson:
\begin{equation}
M_Z = {1 \over 2} \sqrt{g^2+g'^2} v .
\end{equation}

Now, do we obtain the electric charge non-conservation or
dequantization in the fermion sector, in analogy with the result of
Section 3? Unfortunately, the answer is: no. The reason is this: the
calculation of the Weinberg angle in this model (denoted by
$\theta_1$) shows that this angle is {\em exactly equal} to the
Weinberg angle of the Standard Model:                                 
\begin{equation}
\sin^2 \theta_1 = \sin^2 \theta.
\end{equation}
Note that this exact equality has been obtained without assuming $m^2$
or $v_1^2$ to be small (but, of course, assuming that the condition
of photon masslessness, Eq. \ref{41} holds.)
From this equality it follows that the fermion electromagnetic current
 in this
model remains exactly the same as in the Standard Model:
\begin{equation}
j_{\mu}= e(-\bar{e} \gamma_{\mu}e + {2 \over 3} \bar{u} \gamma_{\mu}u
- {1 \over 3} \bar{d} \gamma_{\mu}d).
\end{equation}
In other words any effects of the electric charge non-conservation or
dequantization are absent {\em in the fermion sector}. Here we would
like to stress an essential point: the absence of these effects in the
fermion sector {\em does not mean} that they are absent altogether.
One should not forget that giving the vacuum expectation to the
charged scalar field $\phi_1$ leads to the electric charge non-
conservation {\em in the scalar sector}. However, from the
phenomenological point of view, these effects are much harder to
observe. Such effects would be similar to those arising in a model
with charged scalar field but without the $m^2$ term. Models of such
type have been considered in the literature before and we do not
intend to go into details here.

To conclude this Section, we see that in the context of the present
model, vanishing of the photon mass leads to vanishing effects of
charge non-conservation and charge dequantization {\em in the fermion
sector} (but not {\em in the scalar sector}
).

\section{Hybrid model with a scalar doublet}

In the previous section we have considered the model with a hard mass
for the field B and the scalar {\em singlet} violating the
electromagnetic U(1) symmetry. In this Section, let us change the
singlet into the scalar {\em doublet}, again violating U(1) symmetry;
the rest of the model will be the same. Thus, the lagrangian of our
new model reads:
\begin{equation}
{\cal L}_2 = {\cal L}_0  +  {1 \over 2}m^2B^2 + 
|(\partial_{\mu} -ig {\tau^a \over 2} A^a_{\mu} -i {g'
\over 2} (1 + \epsilon_2)B_{\mu}) \phi_2|^2 + P(\phi_2 , \phi). 
\end{equation}
where the electric charges of the scalar doublet are:

\begin{equation}
 Q(\phi_2) =\left( \begin{array}{c}
1+ {\epsilon_2 \over 2} \\ {\epsilon_2 \over 2}
\end{array} \right). 
\end{equation}

We break the electromagnetic symmetry by assuming
\begin{equation}
\langle \phi_2 \rangle = {1 \over \sqrt{2}} \left( 
\begin{array}{c}
0 \\ v_2
\end{array} \right). 
\end{equation}
After the spontaneous breakdown of symmetry the mass 
matrix of neutral gauge fields  takes the form:
\begin{equation}
M^{(2)} = M^0 + \Delta M^{(2)} = \left( \begin{array}{cc}
M^0_{33} + {1 \over 4} g^2 v^2_2 
& M^0_{34} - {1 \over 4}gg'(1 + \epsilon_2)v^2_2 \\
M^0_{43} - {1 \over 4}gg'(1 + \epsilon_2)v^2_2 
 & M^0_{44} + m^2 +{1 \over 4} g'^2 (1 + \epsilon_2)^2 v_2^2 
\end{array}
\right).
\end{equation}
The condition for the photon to be massless is:
\begin{equation}                                 
m^2 + {1 \over 4} \epsilon^2 g'^2 {v^2 v^2_2 \over v^2 + v^2_2} 
= 0
\end{equation}

From now on, we will assume that this condition is satisfied. Then,
the mass of Z-boson is given by:
\begin{equation} 
M_Z^2 = {1 \over 4}(g^2 + g'^2)v^2 + {1 \over 4}g^2 v_2^2 + 
{1 \over 4}g'^2(1 + \epsilon_2)^2 v_2^2 + m^2.
\end{equation}
For the Weinberg angle we obtain, neglecting the terms of the order
of $\epsilon_2^2$:
\begin{equation} 
\sin^2 \theta_2 = {g'^2(v^2 + (1+ 2\epsilon_2)v_2^2) 
\over (g^2 + g'^2)v^2 + (g^2 + g'^2(1 + 2\epsilon_2 ))v_2^2}
\end{equation}

Assuming that the vacuum expectation of the second doublet is much
smaller than that of the Higgs doublet, we can write down a simpler
expression:
\begin{equation}
\sin^2\theta_2=\sin^2\theta(1+2\epsilon_2\cos^2\theta{v_2^2 
\over v^2}),
\end{equation}
where $\theta$ is the Weinberg angle {\em of the Standard Model}.
As before, the electromagnetic interaction is given by Eq. \ref{12}--
\ref{14}
in which
$\sin \theta$ has to be substituted by $\sin \theta_2$:
\begin{eqnarray}
{\cal L}^{em}_2 &=& {\cal L}_{2l}^{em} + {\cal L}_{2q}^{em} \\
{\cal L}_{2l}^{em} &=& A_{\mu}[{1 \over 2}(g\sin\theta_2 - 
g'\cos\theta_2)
\bar{\nu}_{L}\gamma^{\mu}{\nu}_L
-{1 \over 2}(g\sin\theta_2 + g'\cos\theta_2) 
\bar{e}_{L}\gamma^{\mu}{e}_L
\nonumber\\
&& - g'\cos\theta_2 \bar{e}_{R}\gamma^{\mu}{e}_R] \\
{\cal L}_{2q}^{em} &=& A_{\mu}[{1 \over 2}
g\sin\theta_2(\bar{u}_{L}\gamma^{\mu}{u}_L -
\bar{d}_{L}\gamma^{\mu}{d}_L) + g'\cos\theta_2({1 \over
6}\bar{u}_{L}\gamma^{\mu}{u}_L +{1 \over
6}\bar{d}_{L}\gamma^{\mu}{d}_L + \nonumber\\
&& {2 \over
3}\bar{u}_{R}\gamma^{\mu}{u}_R - {1 \over
3}\bar{d}_{R}\gamma^{\mu}{d}_R)]
\end{eqnarray}                                 
We see that the charge dequantization and charge non-conservation
effects are controlled by the parameter
\begin{equation}
\delta = g \sin \theta_2 -g'\cos \theta_2.
\end{equation}
This parameter measures the deviation of our
theory from the Standard Model (in the latter $g \sin \theta -g'\cos
\theta =0$). Up to the terms of the order of ${v_2^2 \over v^2}$ we
have:
\begin{equation}
\delta=e\epsilon_2{v^2_2 \over v^2}.
\end{equation}
In terms of $\delta$ we can conveniently express the dequantized
lepton and quark charges.

The neutrino charge is:
\begin{equation}
Q_{\nu}= {1 \over 4} \delta.
\end{equation}
The axial electron charge is equal to:
\begin{equation}
Q^5_e= - {1 \over 4} \delta.
\end{equation}
Our normalization is such that the vector electron charge should
 coincide exactly with $-e$, without any corrections:
\begin{equation}
Q_e= -e.
\end{equation}
The vector ($Q_u$) and the axial ($Q_u^5$) charges of u-quark are
given by:
\begin{eqnarray}
Q_u &=& {2 \over 3}e + {1 \over 12}\delta \\
Q_u^5 &=& {1 \over 4} \delta.
\end{eqnarray}
The charges of d-quark are equal to:
\begin{eqnarray}
Q_d &=& -{1 \over 3}e - {1 \over 6} \delta \\
Q_d^5 &=& -{1 \over 4} \delta.
\end{eqnarray}
Consequently, the vector charge of the neutron is:
\begin{equation}
Q_n = Q_u + 2Q_d = -{1 \over 4} \delta.
\end{equation}
The vector charge of the proton equals
\begin{equation}
Q_p = 2Q_u + Q_d = e.
\end{equation}
Therefore, although the electric charge is dequantized in this model,
nevertheless the following relations between the fermion charges hold
true:
\begin{equation}
Q_n + Q_{\nu} =0 ; \;\;\; Q_p + Q_e =0.
\end{equation}
From various experiments testing the validity of electric charge
quantization we can infer the following upper bounds on the parameter
$\delta$.
From the upper bound (\cite{brf,bc}) on the (electron) neutrino 
charge:
\begin{equation}
 \delta < 4 \times 10^{-13}  or 4 \times 10^{-17}.
\end{equation}
From the constraint (\cite{n}) on the neutron electric charge:
\begin{equation}
 \delta < 4 \times 10^{-21}  .
\end{equation}
From the tests (\cite{ep}) of the neutrality of atoms:
\begin{equation}
 \delta < 4 \times 10^{-18} .
\end{equation}

\section{Conclusion and outlook}
To conclude, we have studied a series of hybrid $SU(2) \times U(1)$
models in which the symmetry is broken down not only spontaneously
 but also
explicitely by adding a hard mass term for the $U(1)$ field in 
the lagrangian.  We study the issue of electric charge 
nonconservation
and dequantization in these models. For this purpose we
construct and analyze a series of hybrid models with different
scalar contents. 

 In the minimal hybrid model the electric charge
is not conserved (even though there are no charge violating vacuum
expectation values). The reason is that the mixing angle between
the photon and the Z-boson gets changed as compared with the 
Standard Model, so that the electromagnetic current receives an 
additional non-conserved contribution. The same reason accounts
for the fact that the electric charges become slightly different
from their standard values (i.e., charge dequantization occurs).
In this minimal model  the photon acquires a
non-zero mass which puts a tight limit on the allowed magnitude
of the hard mass term for the $U(1)$ field.

Next, we added to the minimal hybrid model a scalar singlet 
with a non-zero  electric charge. We then assumed that this 
singlet has a non-vanishing vacuum expectation thus violationg
spontaneously the electromagnetic symmetry. We showed that the 
photon mass (at the tree level) receives two contributions: the
 first from the hard mass term of the $U(1)$ field, and the second
from the scalar vacuum expectation. The parameters can be chosen
such that these two terms cancel against each other so that
the photon remains massless (at the tree level). If this is done,
the Weinberg angle is not changed, therefore the
 electric charges of the fermions remain the same and there
is no electric charge non-conservation in the fermion sector
(but in the scalar sector the electric charge  is not conserved).

Finally, we presented a hybrid model with an extra scalar doublet
(in addition to the Higgs doublet of the Standard Model). Again, 
we assumed that the doublet spontaneously violates the 
electromagnetic symmetry. In analogy with the previous model,
the photon mass again consists of two terms: one due to the $U(1)$
hard mass term and the other due to the new scalar doublet
vacuum average. Also, we can arrange for these two terms
to cancel. 

However, here starts the difference with the scalar-singlet model
and an interesting consequence arises. Although the photon mass
is zero (at the tree level),
the Weinberg angle {\em does \/} get modified, so that the electric 
charges of the fermions become {\em dequantized \/} and, moreover, the 
electromagnetic
current (defined as the current interacting with the electromagnetic
field) is {\em no longer conserved \/}.

From the results of the experiments measuring the electric charges
of the neutron and atoms, and the astrophysical limits on the 
neutrino electric charge we have derived upper bounds on the parameter
that governs the effects of charge dequantization and 
non-conservation in our model.

The next important step would be to consider the hybrid models
beyond the tree level and address the problem of renormalizability
of such models.

The authors are grateful to R.Foot and R.Volkas for stimulating
discussions.

This work was supported in part by the Australian Research Council.

\end{document}